\documentclass[11pt]{article}
\usepackage{times}
\usepackage{fullpage}
\usepackage{amsmath}
\usepackage{amssymb}
\usepackage{amsfonts}
\usepackage{epsfig}
\usepackage[all,poly]{xy}

\newcounter{algoctr}

\newif\ifnotesw\noteswtrue% T to show box & marginal notes; F suppresses.
   {\ifnotesw\marginpar[\hfill\(\top\)]{\(\top\)}\fi}%
   {\ifnotesw\marginpar[\hfill\(\bot\)]{\(\bot\)}\fi}

\newcommand{\mnote}[1]%
    {\ifnotesw\marginpar%
        [{\scriptsize\begin{minipage}[t]{\marginparwidth}
        \raggedleft#1%
                        \end{minipage}}]%
        {\scriptsize\begin{minipage}[t]{\marginparwidth}
        \raggedright#1%
                        \end{minipage}}%
    \fi}
\newcommand{\ignore}[1]{}

\newcommand{\etal}{{\it et al. }}

\newsavebox{\given}
\savebox{\given}[1em]{\rule[-1.5ex]{.2mm}{4ex}}

\newcommand{\bnum}{\begin{equation}}
\newcommand{\enum}{\end{equation}}

\newcommand{\blackslug}{\rule{7pt}{7pt}}

\newcommand{\real}{\ifmmode {\rm R} \else ${\rm R}$ \fi}
\newcommand{\nat}{\ifmmode {\rm N} \else ${\rm N}$  \fi}
\newcommand{\tot}{\ifmmode {\cal T} \else ${\cal T}$ \fi}
\newcommand{\sigstar}{\ifmmode \Sigma^{\ast} \else $\Sigma^{\ast}$ \fi}

\newcommand{\inn}{\ifmmode \in \else $\in$ \fi}
\renewcommand{\phi}{\ifmmode \varphi \else $\varphi$ \fi}
\renewcommand{\le}{\ifmmode \leq \else $\leq$ \fi}
\renewcommand{\ge}{\ifmmode \geq \else $\geq$ \fi}
\renewcommand{\ne}{\ifmmode \neq \else $\neq$ \fi}
\newcommand{\lt}{\ifmmode < \else $<$ \fi}
\newcommand{\gt}{\ifmmode > \else $>$ \fi}
\newcommand{\eq}{\ifmmode = \else $=$ \fi}
\newcommand{\half}{\ifmmode \frac{1}{2} \else $\frac{1}{2}$ \fi}
\newcommand{\oneovern}{\ifmmode \frac{1}{n} \else $\frac{1}{n}$ \fi}

\newcommand{\ra}{\ifmmode \rightarrow \else $\rightarrow$ \fi}

\renewcommand{\notin}{\ifmmode \not\in \else $\not\in$ \fi}
\newlength{\thislabel}
\newcommand{\labsize}[1]{\settowidth{\thislabel}{#1}}

\newcommand{\lip}{\langle}
\newcommand{\rip}{\rangle}

\def\Complex{\mathbb C}
\def\Int{\mathbb Z}
\def\Path{\mathbb P}
\def\Real{\mathbb R}

\newcommand{\bra}[1]{\lip #1 |}
\newcommand{\ket}[1]{| #1 \rip}
\newcommand{\braket}[2]{\lip #1 | #2 \rip}
\newcommand{\tderiv}[1]{\frac{\mathsf{d}}{\mathsf{d}t} #1}
\newcommand{\dt}{\mathsf{d}t}
\newcommand{\hlf}{\mbox{\small $\frac{1}{2}$}}
\title{{\bf One-dimensional continuous-time quantum walks}}	%\\ and the Laplace transform}}
\author{
{\sc Daniel ben-Avraham}\footnote{Dept. Physics, Clarkson University. Email: qd00@clarkson.edu. Supported by NSF grant NSF PHY-0140094.} 
\and 
{\sc Erik M. Bollt}\footnote{Dept. Mathematics and Computer Science, and Dept. Physics, Clarkson University. Email: bolltem@clarkson.edu. Supported by NSF grant NSF DMS-0404778.}
\and 
{\sc Christino Tamon}\footnote{Dept. Mathematics and Computer Science, and Center for Quantum Device Technology, Clarkson University. Email: tino@clarkson.edu. Supported by NSF grant DMR-0121146.
}
}
\date{\today}
\begin{document}
\bibliographystyle{plain}
\maketitle

\begin{abstract}
We survey the equations of continuous-time quantum walks on simple one-dimensional lattices, 
which include the finite and infinite lines and the finite cycle,
and compare them with the classical continuous-time Markov chains.
The focus of our expository article is on analyzing these processes using the Laplace transform 
on the stochastic recurrences.
The resulting time evolution equations, classical versus quantum, are strikingly similar in form, 
although dissimilar in behavior. 
%We compare these with analyses done using spectral methods. % (Fourier transform).
We also provide comparisons with analyses performed using spectral methods. % (Fourier transform).
\end{abstract}

\section{Introduction}

The theory of Markov chains on countable structures is an important area in mathematics and physics. 
A quantum analogue of continuous-time Markov chains on the infinite line is well-studied in physics
(for example, it can be found in \cite{fls}, Chapters 13 and 16). 
More recently, it was studied by Aharonov \etal \cite{adz} and by Farhi and Gutmann \cite{fg}. The latter work placed the
problem in the context of quantum algorithms for search problems on graphs. 
Here the symmetric stochastic matrix of the graph is viewed as a Hamiltonian of the quantum process. 
Using Schr\"{o}dinger's equation with this Hamiltonian, we obtain a quantum walk on the underlying graph, 
instead of a classical random walk.

Recent works on continuous-time quantum walks on finite graphs include the analyses of mixing and hitting
times on the $n$-cube \cite{mr,k}, of mixing times on circulant graphs and Cayley graphs of the symmetric group \cite{abtw,gw}, 
and of hitting times on path-like graphs \cite{cfg,ccdfgs}. Most of these are structural results based on spectral analysis 
of the underlying graphs, such as the $n$-cube, circulant and Cayley graphs, and (weighted) paths.
For example, Moore and Russell \cite{mr} proved that the mixing time of a quantum walk on the $n$-cube is
asymptotically faster than a classical random walk; Kempe \cite{k} proved that the hitting time for vertices
on opposite ends of the $n$-cube is exponentially faster than in a classical random walk. Ahmadi {\em et al.} \cite{abtw} 
and Gerhardt and Watrous \cite{gw} proved that circulants and the Cayley graph of the symmetric group lack the uniform
mixing property found in classical random walks.

A recent work of Childs \etal \cite{ccdfgs} gave intriguing evidence that continuous-time quantum walk is a powerful 
method for designing new quantum algorithms. They analyzed diffusion processes on one-dimensional structures 
(finite path and infinite line) using spectral methods. % (e.g., Green's function). 
Another work by Childs and Goldstone \cite{cg03} explored the application of continuous-time quantum walks to perform
Grover search on spatial lattices.

There is an alternate theory of {\em discrete} quantum walks on graphs, which we will not discuss here.
This alternate model was studied in Aharonov {\em et al.} \cite{aakv} and Ambainis {\em et al.} \cite{abnvw}, but had appeared
earlier in work by Meyer \cite{meyer}. The work by Ambainis \etal \cite{abnvw} had also focused on one-dimensional lattices. 
Recently, Ambainis \cite{ambainis} developed an optimal (discrete) quantum walk algorithm for the fundamental problem of 
Element Distinctness. This offers another idea for developing quantum algorithms.

We survey and (re)derive equations for the continuous-time classical and quantum walks on 
one-dimensional lattices using the Laplace transform that works directly with the recurrences.
The Laplace transform is a well-known tool in stochastic processes (see \cite{chung}) and
it might offer a useful alternative to the Fourier transform in certain settings.

\subsection{Stochastic walks on graphs}

Let $G=(V,E)$ be a simple (no self-loops), countable, undirected graph with adjacency matrix $A$.
Let $D$ be a diagonal matrix whose $j$-th entry is the degree of the $j$-th vertex of $G$.
The Laplacian of $G$ is defined as $H = A - D$.
Suppose that $P(t)$ is a time-dependent probability distribution of a stochastic (particle) process on $G$.
The classical evolution of the continuous-time walk is given by the Kolmogorov equation
\begin{equation} \label{eqn:kolmogorov}
%\tderiv{P(t)} = H P(t).
P'(t) = H P(t).
\end{equation}
The solution this equation, modulo some conditions, is $P(t) = e^{tH}P(0)$, 
which can be solved by diagonalizing the symmetric matrix $H$. 
This {\em spectral} approach requires full knowledge of the spectrum of $H$.

A quantum analogue of this classical walk uses the Schr\"{o}dinger equation
in place of the Kolmogorov equation.
Let $\psi: V(G) \rightarrow \Complex$ be the time-independent amplitude of the quantum process on $G$.
Then, the wave evolution is
\begin{equation} \label{eqn:schrodinger}
%i\hslash \tderiv{\psi(t)} = H \psi(t).
%i\hslash \psi'(t) = H \psi(t).
i\hslash \tderiv{\psi(t)} = H \psi(t).
\end{equation}
Assuming $\hslash = 1$ for simplicity, the solution of this equation is
$\psi(t) = e^{-itH}\psi(0)$, which, again, is solvable via spectral techniques.
%Note that $e^{-itH}$ is unitary, since $H$ is Hermitian.
The classical behavior of this quantum process is given by the probability distribution
$P(t)$ whose $j$-th entry is $P_{j}(t) = |\psi_{j}(t)|^2$, where $\psi_{j}(t)=  \braket{j}{\psi(t)}$. 
The {\em average probability} of vertex $j$ is defined as 
$\overline{P}(j) = \lim_{T \rightarrow \infty} \frac{1}{T} \int_{0}^{T} P_{j}(t) \mathsf{d}t$ \
(see \cite{aakv}). 

%\ignore{
\begin{figure}[h]
\hrule
\vspace{.25in}
\begin{center}
\setlength{\unitlength}{0.5cm}
\begin{picture}(21,3)
\thicklines
\put(-1,2){\makebox(0,0){\ldots}}
\put(1,1){\circle*{0.25}}
\put(1,1){\line(1,0){1}}
\put(1,1){\vector(-1,0){1.5}}
\put(1,2){\makebox(0,0){-2}}
\put(2,1){\circle*{0.25}}
\put(2,1){\line(1,0){1}}
\put(2,2){\makebox(0,0){-1}}
\put(3,1){\circle*{0.25}}
\put(3,1){\line(1,0){1}}
\put(3,2){\makebox(0,0){0}}
\put(4,1){\circle*{0.25}}
\put(4,1){\line(1,0){1}}
\put(4,2){\makebox(0,0){1}}
\put(5,1){\circle*{0.25}}
\put(5,1){\line(1,0){1}}
\put(5,2){\makebox(0,0){2}}
\put(5,1){\vector(1,0){1.5}}
\put(6,2){\makebox(0,0){\ldots}}

\put(9,1){\circle*{0.25}}
\put(9,1){\line(1,0){1}}
\put(9,2){\makebox(0,0){0}}
\put(10,1){\circle*{0.25}}
\put(10,1){\line(1,0){1}}
\put(10,2){\makebox(0,0){1}}
\put(11,1){\circle*{0.25}}
\put(11,1){\line(1,0){1}}
\put(11,2){\makebox(0,0){2}}
\put(12,1){\circle*{0.25}}
\put(12,1){\line(1,0){1}}
\put(12,2){\makebox(0,0){3}}
\put(13,1){\circle*{0.25}}
\put(13,2){\makebox(0,0){4}}

\put(17,1){\circle*{0.25}}
\put(16.5,1){\makebox(0,0){0}}
\put(17,3){\circle*{0.25}}
\put(16.5,3){\makebox(0,0){1}}
\put(19,1){\circle*{0.25}}
\put(19.5,1){\makebox(0,0){2}}
\put(19,3){\circle*{0.25}}
\put(19.5,3){\makebox(0,0){3}}
\put(17,1){\line(1,0){2}}
\put(17,1){\line(0,1){2}}
\put(19,3){\line(-1,0){2}}
\put(19,1){\line(0,1){2}}
\end{picture}
\vspace{.05in}
\hrule

%\[\begin{xy} /r10mm/:
% ,+(0.0,0.5),{\xypolygon8"C"{~={0}\dir{*}}}
%\end{xy}\]
\caption{Examples of some one-dimensional lattices.
From left to right: $\Int$, $\Path_{4}$, $\Int_{4}$.}
\label{figure:lattice}
\end{center}
\end{figure}
%}

\vspace{.01in}

\begin{figure}[h]
\begin{center}
\begin{tabular}{|c|c|c|} \hline
{\bf Graph} &	{\bf Classical walk }	& 	{\bf Quantum walk } \\ 
	&	$P_{j}(t)=\mbox{ probability on vertex $j$ at time $t$}$		&	$\psi_{j}(t)=\mbox{ amplitude on vertex $j$ at time $t$}$		\\ \hline
	&				&				\\
$\Int$	& 	$e^{-t}I_{|j|}(t)$ 	& 	$(-i)^{|j|}J_{|j|}(t)$	\\ %\hline
	&				&				\\
$\Int_{N}$ 
	& 	$\sum\limits_{\alpha \equiv \pm j(\mbox{\scriptsize mod}N)} e^{-t}I_{\alpha}(t)$
	& 	$\sum\limits_{\alpha \equiv \pm j(\mbox{\scriptsize mod}N)} (-i)^{\alpha} J_{\alpha}(t)$ \\ %\hline
	&				&				\\
$\Path_{N}$	
	& 	$\sum\limits_{\alpha \equiv \pm j(\mbox{\scriptsize mod }2N)} e^{-t}I_{\alpha}(t)$
	& 	$\sum\limits_{\alpha \equiv \pm j(\mbox{\scriptsize mod }2N)} (-i)^{\alpha} J_{\alpha}(t)$ \\
	&				&				\\ \hline
\end{tabular}
\end{center}
\caption{The equations of the continuous-time classical versus quantum walks on the infinite line, finite cycle, and the finite line,
assuming the particle starts at position $0$.}
\label{figure:equations}
\end{figure}

\vspace{.01in}

\begin{figure}[h]
\begin{center}
\epsfig{file=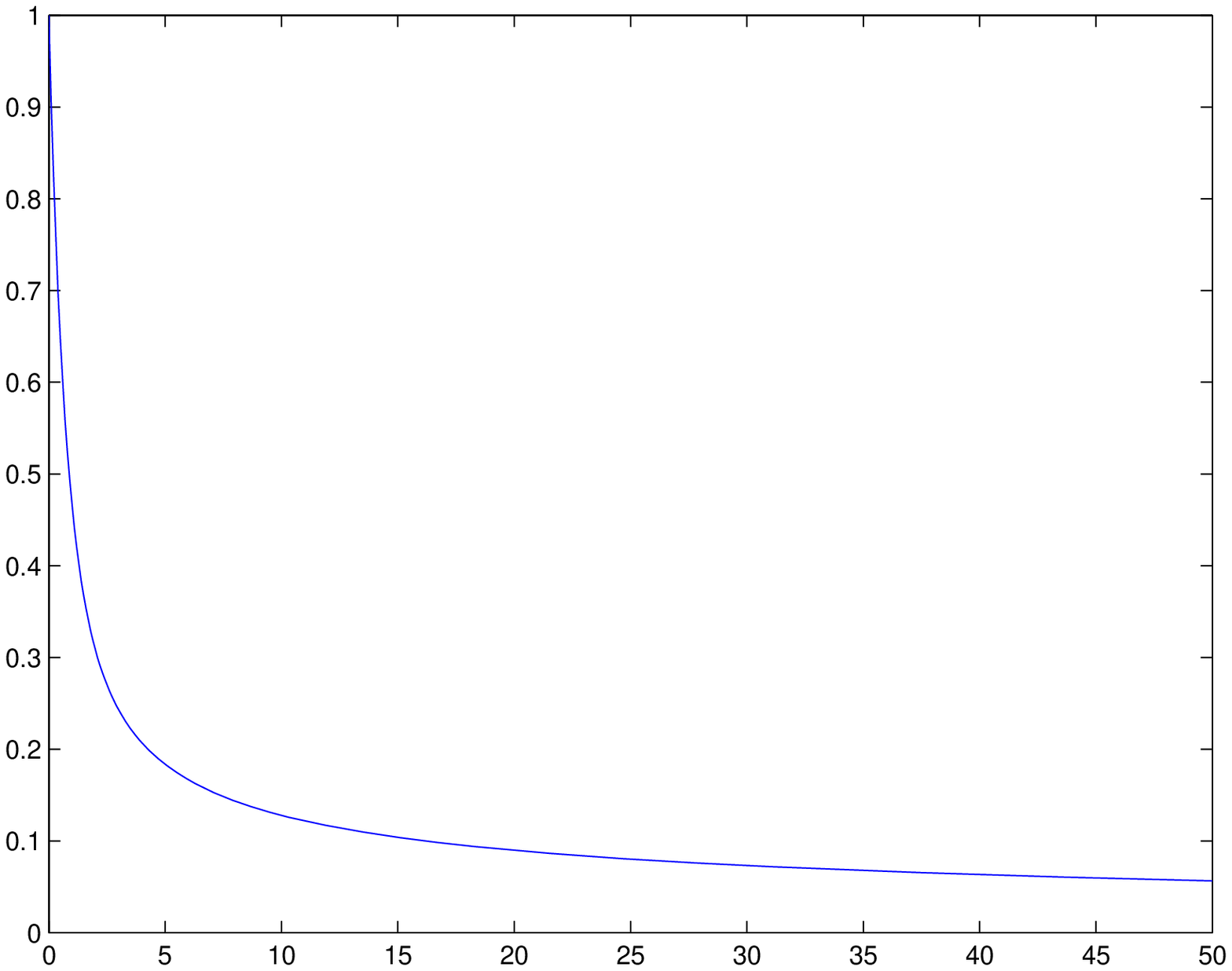, height=6cm, width=5cm}
\hspace{1in}
\epsfig{file=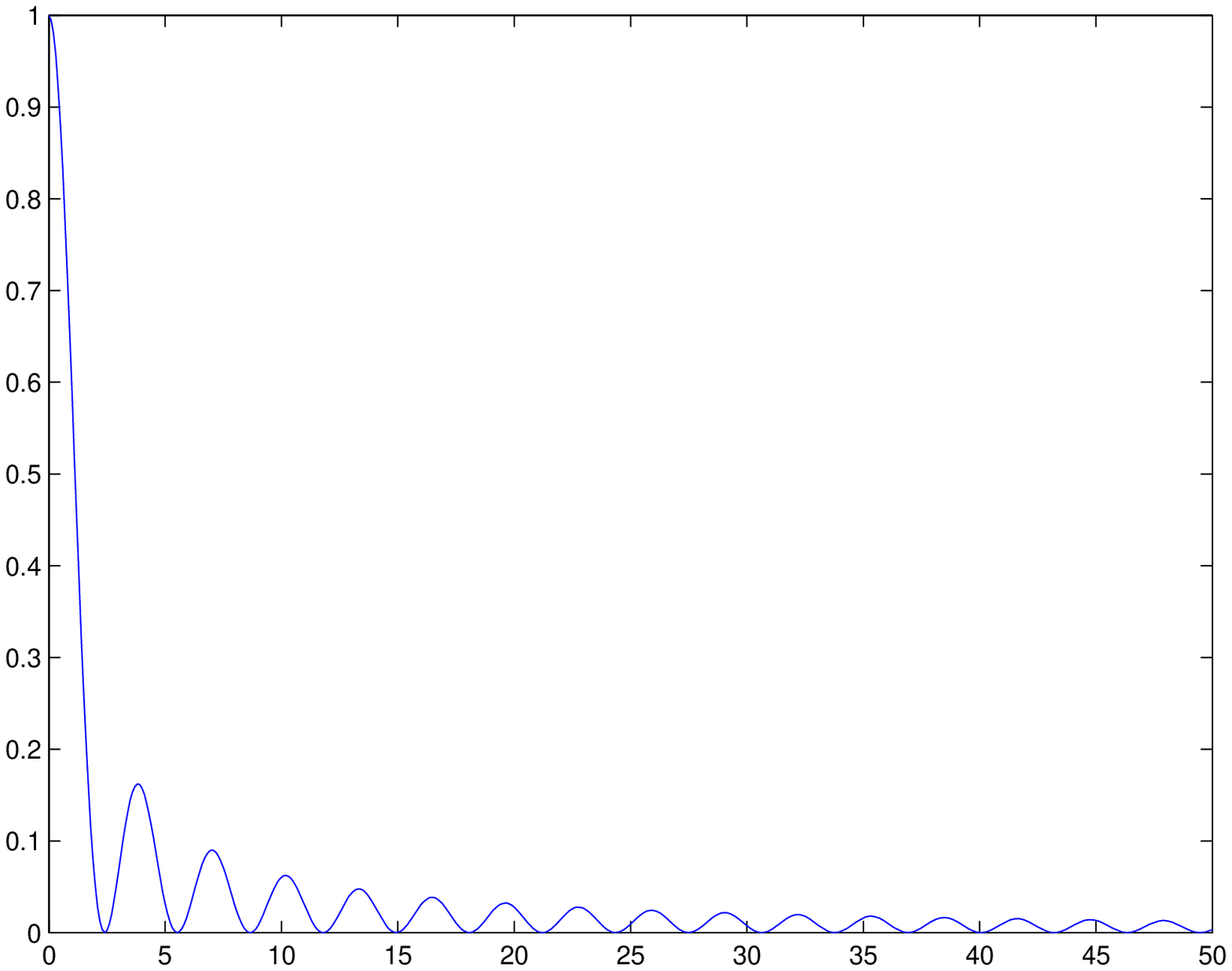, height=6cm, width=5cm}
\caption{Stochastic walks on the infinite line $\Int$: 
(a) plot of $P_{0}(t)$ in the continuous-time random walks for $t \in [0,50]$.
(b) plot of $|\psi_{0}(t)|^{2}$ in a continuous-time quantum walk for $t \in [0,50]$.
Both processes exhibit exponential decay, but with the quantum walk showing an oscillatory behavior.}
\label{figure:plot1}
\end{center}
\end{figure}

\begin{figure}[h]
\epsfig{file=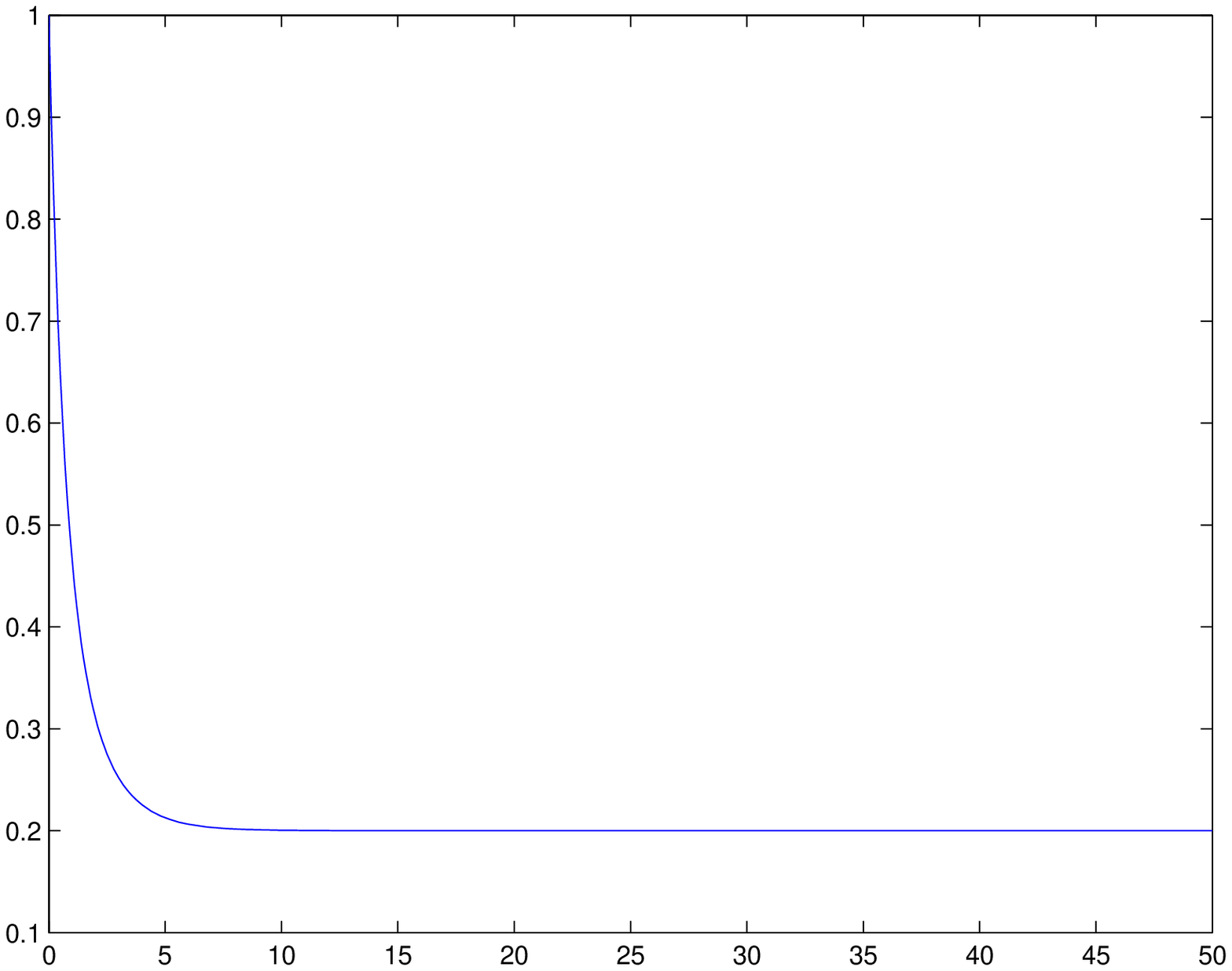, height=6cm, width=5cm}
\hspace{.5in}
\epsfig{file=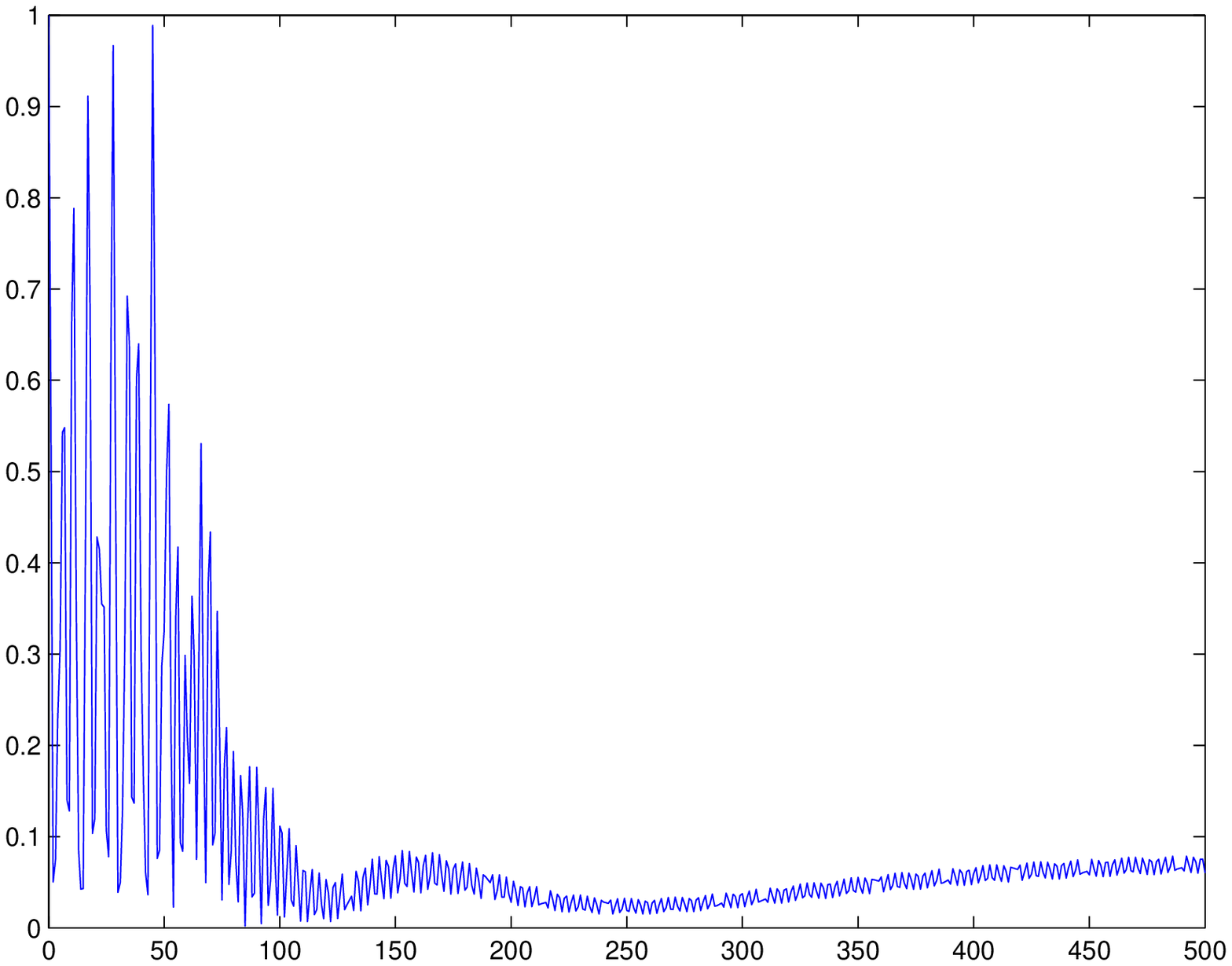, height=6cm, width=10cm}
\caption{Stochastic walks on the finite cycle $\Int_{5}$, each approximated using $21$ terms:
(a) plot of $P_{0}(t)$ in the continuous-time random walks for $t \in [0,50]$.
(b) plot of $|\psi_{0}(t)|^{2}$ in the continuous-time quantum walk for $t \in [0,500]$.
The classical walk settles quickly to $1/5$, while the quantum walk exhibit a short-term chaotic
behavior and a long-term oscillatory behavior below $0.1$.}
\label{figure:plot2}
\end{figure}

The table in Figure \ref{figure:equations} shows the known equations for continuous-time stochastic walks on 
the infinite (integer) line $\Int = \{\ldots, -2,-1,0,1,2,\ldots\}$, 
the finite cycle $\Int_{N} = \{0,\ldots,N-1\}$ on $N$ vertices, 
and the finite path $\Path_{N} = \{0,\ldots,N\}$ on $N+1$ vertices, in terms of the
two kinds of Bessel functions $I(\cdot)$ and $J(\cdot)$.
We assume here that the particle is initially at $0$.
%Our goal is to rederive these equations using arguments based on Laplace transform.
The plots in Figures \ref{figure:plot1} and \ref{figure:plot2} show the dissimilar behavior 
of the classical versus quantum walks.

\subsection{Laplace transform}

The Laplace transform of a time-dependent function $P(t)$, denoted $\hat{P}(s) = \mathcal{L}\{P(t)\}$, is defined as
\begin{equation}
%\hat{P}(s) \dotdef 
\mathcal{L}\{P(t)\} = \int_{0}^{\infty} e^{-st}P(t) \ \dt.
\end{equation}
The only basic properties of the Laplace transform which we will need are (see \cite{as}):
\begin{itemize}
\item Linearity:
$\mathcal{L}\{aP(t) + bQ(t)\} = a\hat{P}(s) + b\hat{Q}(s)$
\item Derivative:
$\mathcal{L}\{P'(t)\} = s\hat{P}(s) - P(0)$
\item Shifting:
$\mathcal{L}\{e^{at}P(t)\} = \hat{P}(s-a)$
\end{itemize}
The relevant Inverse Laplace transform involving the Bessel %$I_{\nu}(t)$, $J_{\nu}(t)$ 
functions are (for $\nu > -1$):
\begin{equation} \label{eqn:bessel-I}
\hat{P}(s) = \frac{(s - \sqrt{s^2 - a^2})^{\nu}}{\sqrt{s^2 - a^2}} \ \ \
\Longleftrightarrow \ \ \
P(t) = a^{\nu}I_{\nu}(at) \ \ \ \mbox{ (Eqn. 29.3.59 in \cite{as}) } 
\end{equation} 
\begin{equation} \label{eqn:bessel-J}
\hat{P}(s) = \frac{(\sqrt{s^2 + a^2} - s)^{\nu}}{\sqrt{s^2 + a^2}} \ \ \
\Longleftrightarrow \ \ \
P(t) = a^{\nu}J_{\nu}(at) \ \ \ \mbox{ (Eqn. 29.3.56 in \cite{as}) }
\end{equation}

\section{The infinite line}

%\subsection{Classical walk}

\par\noindent{\bf Classical process. }
The Kolmogorov equation for the infinite line is
\begin{equation} \label{eqn:classical-line}
%\tderiv{{P}_{j}} = \frac{1}{2}P_{j-1} - P_{j} + \frac{1}{2}P_{j+1},
P_{j}'(t) = \hlf P_{j-1}(t) - P_{j}(t) + \hlf P_{j+1}(t),
\end{equation}
with initial value $P_{j}(0) = \delta_{0,j}$. 
The Laplace transform of (\ref{eqn:classical-line}) is
\begin{equation}
\hat{P}_{j+1}(s) - 2(s+1)\hat{P}_{j}(s) + \hat{P}_{j-1}(s) = - P_{j}(0). 
\end{equation}
The solution of $q^2 - 2(s+1)q + 1$ is $q_{\pm} = (s+1) \pm \sqrt{(s+1)^2-1}$.
A natural guess of the solution is
\begin{equation}
\hat{P}_{j}(s) = 
	\left\{\begin{array}{ll}
	A q_{+}^{j}	& 	\mbox{ if $j < 0$ } \\
	A q_{-}^{j}	& 	\mbox{ if $j > 0$ }
	\end{array}\right.
\end{equation}
When $j = 0$, we get $A = (1 + s - q_{-})^{-1}$.
Thus, for $j \in \Int$,
\begin{equation}
\hat{P}_{j}(s) 
	= \frac{q_{-}^{|j|}}{(1+s-q_{-})}
	= \frac{((s+1) - \sqrt{(s+1)^2-1})^{|j|}}{\sqrt{(s+1)^2 - 1}}.
\end{equation}
Using the Inverse Laplace transform (\ref{eqn:bessel-I}), after shifting $S=s+1$, we get
\begin{equation}
P_{j}(t) = e^{-t}I_{|j|}(t).
\end{equation}
This is a probability function, since 
$e^{t/2(z + 1/z)} = \sum_{k=-\infty}^{\infty} z^{k}I_{k}(t)$, if $z \neq 0$
(see Eqn. 9.6.33 in \cite{as}).

\vspace{.2in}

%\subsection{Quantum walk}

\par\noindent{\bf Quantum process. }
The Schr\"{o}dinger equation for the infinite line is
\begin{equation} \label{eqn:quantum-line}
%i \tderiv{\psi_{j}} = \frac{1}{2}\psi_{j-1} + \frac{1}{2}\psi_{j+1}.
i \psi_{j}'(t) = \hlf \psi_{j-1}(t) + \hlf \psi_{j+1}(t).
\end{equation}
The Laplace transform of (\ref{eqn:quantum-line}) is
\begin{equation} \label{eqn:qgeneral-line}
\hat{\psi}_{j+1}(s) - 2i(s \hat{\psi}_{j}(s) - \psi_{j}(0)) + \hat{\psi}_{j-1}(s) = 0
\end{equation}
The solutions of $q^2 - 2isq + 1$ are $q_{\pm} = i(s \pm \sqrt{s^2+1})$,
where $q_{+}q_{-} = 1$. A guess for the solution is
\begin{equation}
\hat{\psi}_{j}(s) = 
	\left\{\begin{array}{ll}
	A q_{+}^{j}	& 	\mbox{ if $j \le 0$ } \\
	A q_{-}^{j}	& 	\mbox{ if $j > 0$ }
	\end{array}\right.
\end{equation}
When $j = 0$, we get $A = (s + iq_{-})^{-1}$. Thus,
\begin{equation}
\hat{\psi}_{j}(s) 
	= \frac{q_{-}^{|j|}}{(s + iq_{-})}
	= (-i)^{|j|}\frac{(\sqrt{s^2+1}-s)^{|j|}}{\sqrt{s^2+1}}.
\end{equation}
The Inverse Laplace transform (\ref{eqn:bessel-J}) yields, for $j \in \Int$,
\begin{equation}
\psi_{j}(t) = (-i)^{|j|}J_{|j|}(t),
\end{equation}
This is a probability function, since 
$1 = J_{0}^{2}(z) + 2\sum_{k=1}^{\infty} J_{k}^{2}(z)$
(see Eqn. 9.1.76 in \cite{as}).

\vspace{.2in}

\par\noindent{\bf Spectral analysis. }
Let $H$ be a Hamiltonian defined as $\bra{j}H\ket{k} = \frac{1}{2}$ if $j = k \pm 1$, and $0$ otherwise.
For each $p \in [-\pi,\pi]$, define $\ket{p}$ so that 
\begin{equation}
\braket{j}{p} = \frac{1}{\sqrt{2\pi}} e^{ipj}.
\end{equation}
The eigenvalue equation $H\ket{p} = \lambda_{p}\ket{p}$ has the solution $\lambda_{p} = \cos(p)$.
Thus, the amplitude of position $j$ when the particle starts at position $0$ is
\begin{equation}
\bra{j}e^{-iHt}\ket{0} = \frac{1}{2\pi} \int_{-\pi}^{\pi} e^{ipj} e^{-it\cos(p)} \ \mathsf{d}p
	= (-i)^{j}J_{j}(t) \ \ \ \mbox{ (see Eqn. 9.1.21 in \cite{as})}
\end{equation}
Childs \etal \cite{ccdfgs} gave a more generalized analysis along these lines.

\section{The finite cycle}

%\subsection{Classical walk}

\par\noindent{\bf Classical process. }
If $A$ is the adjacency matrix of the finite cycle, let $H = \frac{1}{2}A - I$ be its Laplacian matrix. 
The Kolmogorov equation for the finite cycle is 
\begin{equation} \label{eqn:classical-cycle}
%\tderiv{P_{j}} = \frac{1}{2}P_{j-1} - P_{j}+ \frac{1}{2}P_{j+1}.
P_{j}'(t) = \hlf P_{j-1}(t) - P_{j}(t) + \hlf P_{j+1}(t).
\end{equation}
Applying the Laplace transform to (\ref{eqn:classical-cycle}), we get
\begin{equation}
%s\hat{P}_{j} - P_j(0) = \frac{1}{2}\hat{P}_{j-1} - \hat{P}_{j} + \frac{1}{2}\hat{P}_{j+1} \ \ \
%\Longleftrightarrow
(s+1)\hat{P}_{j}(s) - P_j(0) = \hlf \hat{P}_{j-1}(s) + \hlf \hat{P}_{j+1}(s).
\end{equation}
For convenience, define the extra condition
%\begin{equation} \label{eqn:minus-one}
$\hat{P}_{-1}(s) = \hat{P}_{N-1}(s) + 2$,
%\end{equation}
so that 
%\begin{equation} \label{eqn:general}
$\hat{P}_{j+1}(s) -2(s+1)\hat{P}_{j}(s) + \hat{P}_{j-1}(s) = 0$
%\end{equation}
holds for $j \in \Int_{N}$. The cycle condition is
%\begin{equation}
$\hat{P}_{N}(s) = \hat{P}_{0}(s)$.
%\end{equation}
We guess the solution to be 
\begin{equation} 
\hat{P}_{j}(s) = A q_{+}^{j} + B q_{-}^{j},
\end{equation}
where $q_{\pm}$ is the solution to $x^2 - 2(s+1)x + 1 = 0$, i.e.,
$q_{\pm} = (s+1) \pm \sqrt{(s+1)^2 - 1}$, with $q_{+}q_{-} = 1$.
Using the cycle condition, we get
\begin{equation} \label{eqn:AB}
A q_{+}^{N} + B q_{-}^{N} = A + B \ \
\Longrightarrow \ \
A(q_{+}^{N}-1) = B(1-q_{-}^{N}) \ \
\Longrightarrow \ \
B = A q_{+}^{N}.
\end{equation}
Using the extra condition
%Equations (\ref{eqn:minus-one}) 
and (\ref{eqn:AB}), we get 
%\begin{equation}
$A = 2((q_{+}-q_{-})(q_{+}^{N}-1))^{-1}$. 
%\ \ \ B = \frac{2q_{+}^{N}}{(q_{+}-q_{-})(q_{+}^{N}-1)}.
%\end{equation}
Thus, for $j \in \Int_{N}$, 
\begin{eqnarray*}
\hat{P}_{j}(s) & = & A q_{+}^{j} + B q_{-}^{j} = A(q_{+}^{j} + q_{+}^{N-j}) \\
	& = & \frac{2}{(q_{+}-q_{-})} \frac{(q_{-}^{j} + q_{-}^{N-j})}{(1 - q_{-}^{N})} 
		= \frac{2}{(q_{+}-q_{-})} \sum_{k=0}^{\infty} \left(q_{-}^{kN+j} + q_{-}^{(k+1)N-j}\right) \\
	& = & \sum_{k=0}^{\infty} \left[\frac{((s+1)-\sqrt{(s+1)^2-1})^{kN+j}}{\sqrt{(s+1)^2-1}} +
			\frac{((s+1)-\sqrt{(s+1)^2-1})^{(k+1)N-j}}{\sqrt{(s+1)^2-1}}\right].
\end{eqnarray*}
The Inverse Laplace transform (\ref{eqn:bessel-I}), after shifting, %i.e., ${\cal L}[e^{-at}F(t)] = f(s+a)$, 
yields, for $j \in \Int_{N}$,
\begin{equation}
P_{j}(t) = \sum_{k=0}^{\infty} e^{-t}\left[I_{kN+j}(t) + I_{(k+1)N-j}(t)\right]
	= \sum_{\alpha \equiv \pm j (\mbox{\scriptsize mod} N)} e^{-t}I_{\alpha}(t).
\end{equation}

\vspace{.2in}

%\subsection{Quantum walk}

\par\noindent{\bf Quantum process. }
Since the finite cycle is a regular graph, instead of the Laplacian, we use the adjacency matrix directly.
In a continuous-time quantum walk, this simply introduces an irrelevant phase factor in the final expression.
The Schr\"{o}dinger equation, in this case, is
\begin{equation} \label{eqn:quantum-cycle}
%i \tderiv{\psi_{j}} = \frac{1}{2}\psi_{j-1} + \frac{1}{2}\psi_{j+1}.
i \psi_{j}'(t) = \hlf \psi_{j-1}(t) + \hlf \psi_{j+1}(t).
\end{equation}
The Laplace transform of (\ref{eqn:quantum-cycle}) is
\begin{equation} \label{eqn:qgeneral}
\hat{\psi}_{j+1}(s) - 2i(s \hat{\psi}_{j}(s) - \psi_{j}(0)) + \hat{\psi}_{j-1}(s) = 0
\end{equation}
The cycle boundary condition is
%\begin{equation} \label{eqn:qcircle-boundary}
$\hat{\psi}_{N}(s) = \hat{\psi}_{0}(s)$.
%\end{equation}
For convenience, define
\begin{equation} \label{eqn:qminus-one}
\hat{\psi}_{-1}(s) = \hat{\psi}_{N-1}(s) = 2i.
\end{equation}
The solutions of $q^2 - 2isq + 1$ are $q_{\pm} = i(s \pm \sqrt{s^2+1})$, with $q_{+}q_{-} = 1$.
A solution guess, for $j \in \Int_{N}$, is
\begin{equation} \label{eqn:qsolution-guess}
\hat{\psi}_{j}(s) = Aq_{+}^{j} + Bq_{-}^{j}.
\end{equation}
The cycle boundary condition yields
%From Equation (\ref{eqn:qcircle-boundary}), 
$B = Aq_{+}^{N}$.
By (\ref{eqn:qminus-one}), we get
%\begin{equation}
$A = 2i((q_{+}-q_{-})(q_{+}^{N} - 1))^{-1}$.
%\ \ \ B = \frac{2i}{(q_{+}-q_{-})(1 - q_{-}^{N})}, \ \ \ .
%\end{equation}
Thus, for $j \in \Int_{N}$,
\begin{eqnarray*}
\hat{\psi}_{j}(s) 
	& = & Aq_{+}^{j} + Bq_{-}^{j} = A(q_{+}^{j} + q_{+}^{N-j}) \\
	& = & \frac{2i}{(q_{+} - q_{-})} \frac{(q_{-}^{j} + q_{-}^{N-j})}{(1 - q_{-}^{N})} 
		= \frac{2i}{(q_{+} - q_{-})} \sum_{k=0}^{\infty} \left(q_{-}^{kN+j} + q_{-}^{(k+1)N-j}\right) \\
	& = & \sum_{k=0}^{\infty} \left[\frac{((-i)(\sqrt{s^2+1}-s))^{kN+j}}{\sqrt{s^2+1}} 
		+ \frac{((-i)(\sqrt{s^2+1}-s)^{(k+1)N-j}}{\sqrt{s^2+1}}\right].
\end{eqnarray*}
The Inverse Laplace transform (\ref{eqn:bessel-J}) gives, for $j \in \Int_{N}$,
\begin{equation}
\psi_{j}(t) = \sum_{k=0}^{\infty} \left[(-i)^{kN+j} J_{kN+j}(t) + (-i)^{(k+1)N-j}J_{(k+1)N-j}(t)\right]
	= \sum_{\alpha \equiv \pm j (\mbox{\scriptsize mod}N)} (-i)^{\alpha} J_{\alpha}(t).
\end{equation}

\vspace{.2in}

\par\noindent{\bf Spectral analysis.}
The normalized adjacency matrix $H$ of $\Int_N$ is the circulant matrix
\begin{equation}
H = \left(\begin{array}{cccccc}
    0 & 1/2 & 0 & \ldots & 0 & 1/2 \\
    1/2 & 0 & 1/2 & \ldots & 0 & 0 \\
    0 & 1/2 & 0 & \ldots & 0 & 0 \\
    \vdots & \vdots & \vdots & \vdots & & \vdots \\
    1/2 & 0 & 0 & \ldots & 1/2 & 0
    \end{array}\right).
\end{equation}
It is well-known that all $N \times N$ circulant matrices are unitarily diagonalized
by the Fourier matrix $F = \frac{1}{\sqrt{N}}V(\omega_{N})$,
where $\omega_{N} = e^{2\pi i/N}$ and $V(\omega_{N})$ is the Vandermonde matrix defined
as $V(\omega_{N})[j,k] = \omega_{N}^{jk}$, for $j,k \in \{0,1,\ldots,N-1\}$.
\ignore{
\begin{equation}
	V(\omega) = \left(\begin{array}{lllll}
			1 & 1 & 1 & \ldots & 1 \\
		    	1 & \omega & \omega^2 & \ldots & \omega^{N-1} \\
			1 & \omega^2 & \omega^4 & \ldots & \omega^{2(N-1)} \\
			\vdots & \vdots & \vdots &  & \vdots \\
			1 & \omega^{N-1} & \omega^{2(N-1)} & \ldots & \omega^{(N-1)^2}
		    \end{array}\right).
\end{equation}
}
The eigenvalues of $H$ are
$\lambda_{j} = \frac{1}{2}(\omega_{N}^{j} + \omega_{N}^{j(N-1)}) = \cos(2\pi j/N)$,
for $j=0,1,\ldots,N-1$.
\ignore{
The wave equation of the continuous-time quantum walk is
\begin{equation}
\psi(t) = e^{-it H} \psi(0) %= \frac{U_t}{n}\sum_{j}\ket{\omega_j} = \frac{1}{n}\sum_{j=0}^{n-1} e^{-i\lambda_{j}t}V_{j} = \frac{1}{n}\sum_{j=0}^{n-1}e^{-it\cos(2\pi j/n)}V_{j}.
\end{equation}
}
Thus, the wave amplitude at vertex $j$ at time $t$ is
\begin{equation}
\psi_j(t) = \frac{1}{N}\sum_{k=0}^{N-1} e^{-it\cos(2\pi k/N)}\omega_{N}^{jk}.
\end{equation}
From earlier analysis, we get the following Bessel equation
\begin{equation}
\frac{1}{N}\sum_{k=0}^{N-1} e^{-it\cos(2\pi k/N)}e^{2\pi i jk/N}
=
\sum_{\alpha \equiv \pm j (\mbox{\scriptsize mod } N)} (-i)^{\alpha} J_{\alpha}(t). 
\end{equation}
It is an open question if there exists a time $t \in \Real^{+}$ such that for all $j \in \Int_{N}$
we have $|\psi_{j}(t)|^{2} = 1/N$, i.e., uniformity is achieved at some time $t$. For $N = 2,3,4$,
it is known that {\em instantaneous exact uniform mixing} is achieved (see \cite{mr,abtw}).

\section{The finite path}

%\subsection{Classical walk}

\par\noindent{\bf Classical process. }
Let $A$ be the {\em normalized} adjacency matrix of the finite path, where $A$ is a stochastic
matrix with the probability transitions are proportional to the degrees of the vertices. 
Let $H = A - I$ be its Laplacian. Then, the Kolmogorov equation, in this case, is
\begin{equation} \label{eqn:classical-path}
%\tderiv{P_{j}} = \frac{1}{2}P_{j-1} - P_{j}+ \frac{1}{2}P_{j+1}, 
P_{j}'(t) = \hlf P_{j-1}(t) - P_{j}(t) + \hlf P_{j+1}(t), 
\end{equation}
for $0 < j < N$, with initial condition $P_{j}(0) = \delta_{j,0}$ and boundary conditions
\begin{equation}
%\tderiv{P_{0}} = P_{1} - P_{0}, \ \ \
P_{0}'(t) = P_{1}(t) - P_{0}(t), \ \ \
%\tderiv{P_{N}} = P_{N-1} - P_{N}.
P_{N}'(t) = P_{N-1}(t) - P_{N}(t).
\end{equation}
The Laplace transform of (\ref{eqn:classical-path}) is
\begin{equation}
\hat{P}_{j+1}(s) - 2(s+1)\hat{P}_{j}(s) + \hat{P}_{j-1}(s) = 0, \ \ \ 0 < j < N,
\end{equation}
and two boundary equations
%\begin{equation} \label{eqn:path-starting-boundary}
$(1+s)\hat{P}_{0}(s) - 1 = \hat{P}_{1}(s)$,
%\end{equation}
and
%\begin{equation} \label{eqn:path-nonstarting-boundary}
$(1+s)\hat{P}_{N}(s) = \hat{P}_{N-1}(s)$.
%\end{equation}
A guess of the solution is
\begin{equation} \label{eqn:path-general-solution}
\hat{P}_{j}(s) = A q_{+}^{j} + B q_{-}^{j}, \ \ \ 0 \le j \le N,
\end{equation}
where $q_{\pm} = (s+1) \pm \sqrt{(s+1)^2 - 1}$.
The boundary equations give
%Starting with Equation (\ref{eqn:path-starting-boundary}), we get
%\begin{equation} 
$B - A = 2/(q_{+} - q_{-})$
%\end{equation}
and
%Using Equation (\ref{eqn:path-nonstarting-boundary}), we obtain 
%\begin{equation}
$A = B q_{-}^{2N}$.
%\end{equation}
Combining these last two equations, we get
\begin{equation}
A = \frac{2}{(q_{+} - q_{-})} \frac{q_{-}^{2N}}{(1 - q_{-}^{2N})}.
%\ \ \ B = \frac{2}{(q_{+} - q_{-})} \left(1 + \frac{q_{-}^{2N}}{(1 - q_{-}^{2N})}\right).
\end{equation}
Thus, for $j = 0,1,\ldots,N$, 
\begin{eqnarray*}
\hat{P}_{j}(s) 
	& = & A q_{+}^{j} + B q_{-}^{j} = A(q_{+}^{j} + q_{+}^{2N-j}) %\\
	= \frac{2}{(q_{+} - q_{-})} \frac{(q_{-}^{j} + q_{-}^{2N-j})}{(1 - q_{-}^{2N})} \\
	& = & \frac{2}{(q_{+} - q_{-})} \sum_{k=0}^{\infty} (q_{-}^{2Nk+j} + q_{-}^{2N(k+1)-j}).
\end{eqnarray*}
The Inverse Laplace transform (\ref{eqn:bessel-I}), after shifting, yields, for $j = 0,1,\ldots,N$,
\begin{equation}
P_{j}(t) = \sum_{k=0}^{\infty} e^{-t}\left[I_{2Nk+j}(t) + I_{2N(k+1)-j}(t)\right]
	= \sum_{\alpha \equiv \pm j (\mbox{\scriptsize mod } 2N)} e^{-t}I_{\alpha}(t).
\end{equation}

\vspace{.2in}

%\subsection{Quantum walk}

\par\noindent{\bf Quantum process. }
The Schr\"{o}dinger equation for the finite path is
\begin{equation} \label{eqn:quantum-path}
%i \tderiv{\psi_{j}} = \frac{1}{2}\psi_{j-1} + \frac{1}{2}\psi_{j+1}, 
i \psi_{j}'(t) = \hlf \psi_{j-1}(t) + \hlf \psi_{j+1}(t), 
\end{equation}
for $0 < j < N$, with initial condition $\psi_{j}(0) = \delta_{0,j}$ and boundary conditions
\begin{equation}
%i \tderiv{\psi_{0}} = \psi_{1}, \ \ \ 
i \psi_{0}'(t) = \psi_{1}(t), \ \ \ 
%i \tderiv{\psi_{N}} = \psi_{N-1}.
i \psi_{N}'(t) = \psi_{N-1}(t).
\end{equation}
The Laplace transform of (\ref{eqn:quantum-path}) is
\begin{equation} \label{eqn:qpath-general}
\hat{\psi}_{j+1}(s) - 2is\hat{\psi}_{j}(s) + \hat{\psi}_{j-1}(s) = 0, \ \ \ 0 < j < N,
\end{equation}
and two boundary equations
%\begin{equation} \label{eqn:qpath-starting-boundary}
$is \hat{\psi}_{0}(s) - i = \hat{\psi}_{1}(s)$,
%\end{equation}
and
%\begin{equation} \label{eqn:qpath-nonstarting-boundary}
$is \hat{\psi}_{N}(s) = \hat{\psi}_{N-1}(s)$.
%\end{equation}
The solutions of $q^2 - 2isq + 1$ are $q_{\pm} = i(s \pm \sqrt{s^2+1})$. 
A guess of the solution is
\begin{equation} \label{eqn:qpath-solution}
\hat{\psi}_{j}(s) = Aq_{+}^{j} + Bq_{-}^{j},  \ \ \ 0 \le j \le N.
\end{equation}
From the boundary equations, we get
%Equation (\ref{eqn:qpath-starting-boundary}), we get 
%\begin{equation}
$B - A = 2i/(q_{+} - q_{-})$
%\end{equation}
and
%Equation (\ref{eqn:qpath-nonstarting-boundary}), we derive
%\begin{equation}
$B = A q_{+}^{2N}$.
%\end{equation}
Thus,
\begin{equation}
A = \frac{2i}{(q_{+}-q_{-})}\frac{q_{-}^{2N}}{(1 - q_{-}^{2N})}.
%\ \ \ B = \frac{2i}{(q_{+}-q_{-})}\frac{1}{(1 - q_{-}^{2N})}.
\end{equation}
For $j = 0,1,\ldots,N$, 
\begin{eqnarray*}
\hat{\psi}_{j}(s) 
	& = & Aq_{+}^{j} + Bq_{-}^{j} = A(q_{+}^{j} + q_{+}^{2N-j}) \\
	& = & \frac{2i}{(q_{+} - q_{-})} \frac{(q_{-}^{j} + q_{-}^{2N-j})}{(1-q_{-}^{2N})} 
		= \frac{2i}{(q_{+} - q_{-})} \sum_{k=0}^{\infty} \left(q_{-}^{2Nk+j} + q_{-}^{2(k+1)N-j}\right) \\
	& = & \sum_{k=0}^{\infty} \left[\frac{((-i)(\sqrt{s^2+1}-s))^{2Nk+j}}{\sqrt{s^2+1}}
		+ \frac{((-i)(\sqrt{s^2+1}-s))^{2(k+1)N-j}}{\sqrt{s^2+1}}\right].
\end{eqnarray*}
The Inverse Laplace transform (\ref{eqn:bessel-J}) yields, for $j = 0,1,\ldots,N$,
\begin{equation} \label{eqn:qpath}
\psi_{j}(t) = \sum_{k=0}^{\infty} \left[(-i)^{2Nk+j} J_{2Nk+j}(t) + (-i)^{2N(k+1)-j}J_{2N(k+1)-j}(t)\right]
	= \sum_{\alpha \equiv \pm j (\mbox{\scriptsize mod }2N)} (-i)^{\alpha} J_{\alpha}(t).
\end{equation}

\vspace{.2in}

\par\noindent{\bf Spectral analysis.}
The spectrum of a path on $n$ vertices is given by Spitzer \cite{spitzer}.
For $j \in \{0,1,\ldots,N\}$, the eigenvalue $\lambda_j$ and its eigenvector $v_j$ are given by
\begin{equation}
\lambda_j = \cos\left(\frac{(j+1)\pi}{N+2}\right), \ \ \
v_j(\ell) = \sqrt{\frac{2}{N+2}}\sin\left(\frac{(j+1)\pi}{N+2}(\ell + 1)\right).
\end{equation}
The probability of measuring vertex $0$ at time $t$ is given by
\begin{equation}
P_{0}(t) %= |\bra{0}e^{-itP_n}\ket{0}|^2 
	= \frac{4}{(N+2)^2} \sum_{j,k} \sin^2\left(\frac{(j+1)\pi}{N+2}\right) \sin^2\left(\frac{(k+1)\pi}{N+2}\right)
		e^{-it(\lambda_j-\lambda_k)}.
\end{equation}
\ignore{
\footnote{The average probability of vertex $j$ is defined as 
$\overline{P}(j) = \lim_{T \rightarrow \infty} \frac{1}{T} \int_{0}^{T} P_{j}(t) \mathsf{d}t$ (see \cite{aakv}).} 
}
Since all eigenvalues are distinct, the {\em average probability}
of measuring the starting vertex $0$ is 
\begin{eqnarray*} 
\overline{P}(0) 
	%& = & \lim_{T \rightarrow \infty} \frac{1}{T}\int_{0}^{T} P_{0}(t) \mathsf{d}t \\ 
	& = & \frac{4}{(N+2)^2} \sum_{j,k} \sin^2\left(\frac{(j+1)\pi}{N+2}\right) \sin^2\left(\frac{(k+1)\pi}{N+2}\right) 
		\lim_{T \rightarrow \infty} \frac{1}{T}\int_{0}^{T} e^{-it(\lambda_j-\lambda_k)} \ \dt \\ 
	& = & \frac{4}{(N+2)^2} \sum_{j} \sin^4\left(\frac{(j+1)\pi}{N+2}\right).
\end{eqnarray*} 
Equating this with (\ref{eqn:qpath}), we obtain a Bessel-like equation:
\begin{equation}
	\lim_{T \rightarrow \infty} \frac{1}{T} \int_{0}^{T} 
	%\sum_{a,b \equiv 0(\mbox{\scriptsize mod }2N)} i^{b-a} J_{a}(t)J_{b}(t) 
	\left|\sum_{a \equiv 0(\mbox{\scriptsize mod }2N)} (-i)^{a} J_{a}(t)\right|^{2} \ \dt
	=
	\frac{4}{(N+2)^{2}} \sum_{k=0}^{N} \sin^{4}\left(\frac{(k+1)\pi}{N+2}\right)
\end{equation}

\section{Conclusions}

This expository survey reviews equations for the continuous-time quantum walks on one-dimensional lattices.
The focus was on analysis based on the Laplace transform which works directly with the stochastic recurrences.
It would be interesting to extend this analysis to higher-dimensional or to regular graph-theoretic settings.
Another interesting direction is to consider lattices with defects and weighted graphs \cite{ccdfgs}. 
%We leave these problems for future investigations.

\section*{Acknowledgments}

We thank the anonymous referees for pointing out several omissions and for helpful comments that improve 
the presentation of this manuscript.

\end{document}